\documentclass[preprint,aps,nofootinbib]{revtex4}

\usepackage{epsfig}
\usepackage{graphicx}
\usepackage{amssymb}
\usepackage{amsmath}
\usepackage{subfigure}

\begin{document}

\title{Unitarity Constraints in the standard model with a singlet scalar field}

\author{Sin Kyu Kang}
\email{skkang@snut.ac.kr}
\vskip 0.5cm
\affiliation{School of Liberal Arts, Seoul-Tech., Seoul 139-743, Korea}

\author{ Jubin Park }
\email{honolov77@gmail.com}
\vskip 0.5cm
\affiliation{School of Liberal Arts, Seoul-Tech., Seoul 139-743, Korea}

\vspace{2cm}

\begin{abstract}
Motivated by the discovery of a new scalar field and amelioration of the electroweak vacuum stability
 ascribed to a singlet scalar field embedded in the standard model (SM), we examine the implication of the perturbative unitarity in the SM with a singlet scalar field.
Taking into account the full contributions to the scattering amplitudes,
we derive unitarity conditions on the scattering matrix which can be translated into
bounds on the masses of the scalar fields.
In the case that the singlet scalar field develops vacuum expectation value (VEV), we get
the upper bound on the singlet scalar mass varying with the mixing between the singlet and Higgs scalars.
On the other hand, the mass of the Higgs scalar can be constrained
by the unitarity condition in the case that the VEV of the singlet scalar is not generated.
Applying the upper bound on the Higgs mass to the scenario of the unitarized Higgs inflation, we discuss how the unitarity condition can constrain the Higgs inflation.
The singlet scalar mass is not constrained by the unitarity itself when we impose $Z_2$ in the model because of no mixing with the Higgs scalar.
But, regarding the singlet scalar field as a cold dark matter candidate, we derive upper bound on the singlet scalar mass by combining the observed relic abundance with the unitarity condition. \\

%


\end{abstract}

\maketitle





\section{Introduction}
\label{sec:1}
The discovery of a new scalar particle has been announced by both the ATLAS and CMS collaborations at the large hadron collider(LHC)~\cite{Aad:2012tfa,Chatrchyan:2012ufa}.
At present, the physical properties of the observed new scalar particle seem to be consistent with the long-sought Higgs boson in the standard model(SM), and its mass has been observed at 126 GeV with a few GeV uncertainty~\cite{ATLAS,CMS}.
Interestingly, such a mass range of the SM-like Higgs can imply that the Higgs potential of the SM develops unstable electroweak vacuum at large field values, depending on the top mass and strong coupling constant with some uncertainties~\cite{EliasMiro:2011aa}. From the theoretical point of view, the measurements of the Higgs mass can provide us with an useful hint about the structure of the theory at the very short distance through the sizable renormalization group(RG) running of the Higgs quartic coupling.

Recently a very simple and economical way to stabilize the electroweak vacuum at the high energy has been proposed by introducing one singlet scalar particle and its relevant couplings~\cite{EliasMiro:2012ay}.
 The existence of a heavy singlet scalar can generate threshold corrections to the quartic Higgs coupling which can help to evade the instability of the vacuum at large field values.
On the other hand, embedding the singlet scalar particle in the SM Lagrangian can not only modify the production and/or decay rates of the Higgs field~\cite{Batell:2012mj,Baek:2012uj} but also supply solutions for dark matter~\cite{Burgess:2000yq,Ponton:2008zv}, baryogenesis via the first order electroweak phase transition ~\cite{Profumo:2007wc} and the unitarity problem of the Higgs inflation~\cite{Giudice:2010ka}.

Motivated by the discovery of a new scalar field and the amelioration of the electroweak vacuum stability
 ascribed to the singlet scalar field embedded in the SM, in this paper, we examine the implication of the perturbative unitarity in the SM extended to contain the singlet scalar particle~\cite{Cynolter:2004cq}.
On top of the SM contributions to the scattering amplitudes, we estimate new contributions generated due to the existence of the singlet scalar, and then
derive some conditions that guarantee the perturbative unitarity of the scattering matrix (S-matrix),
which can be translated into some
bounds on the masses of the scalar fields.
In the case that the singlet scalar field develops vacuum expectation value (VEV), we can get
the upper bound on the singlet scalar mass. Thanks to the mixing between the singlet and Higgs scalars, the unitarity bound on the singlet scalar mass depends on the mixing angle between two
scalar fields. As will be shown, the unitarity bound gets stronger as the mixing angle goes up to
maximal.
On the other hand, the mass of the Higgs scalar can be constrained
by the unitarity condition in the case that the VEV of the singlet scalar is not generated.
The upper bound on the Higgs mass derived from the unitarity of the S-matrix in the SM  is well known
as the so-called Lee-Quigg-Thacker (LQT) bound.
The LQT bound is modified and can appear to be severer in the presence of the singlet scalar field.
Although the upper bound on the Higgs mass we derive is not useful to study low energy phenomenology
due to the measurement of the Higgs mass at the LHC, it can be applied to the scenario of the unitarized Higgs inflation. We will discuss how the unitarity condition can constrain the Higgs inflation.
In the model with $Z_2$ symmetry, the mass of the singlet scalar is not constrained by the unitarity itself because of no mixing with the Higgs scalar.
But, regarding the singlet scalar field as a cold dark matter candidate, we can derive upper bound on the singlet scalar mass by combining the observed relic abundance with the unitarity.\,\footnote{In ~\cite{Cynolter:2004cq}\,, the authors have studied the unitarity conditions in the similar model, but considered only limited cases.
The unitarity bound in the models with two Higgs doublets and a triplet scalar have been studied in \cite{Maalampi:1991fb} and \cite{Aoki:2007ah}, respectively.}

This paper is organized as follows. In Sec. II, we briefly present the extension of the SM containing a singlet scalar model and show how three scalar couplings $(\lambda_H, \lambda_S, \lambda_{HS})$ in the model are related to two physical scalar masses, two mixing angles and VEV. In Sec. III, we derive the unitarity condition on the scattering amplitudes by analyzing the eigenvalues of the S-matrix presented in terms of those scalar couplings.  From the numerical analysis, we show how severe the unitarity conditions can constrain the masses of the scalar fields. In Sec. IV, we discuss about the applications of the unitarity conditions to the unitarized Higgs inflation and the singlet scalar dark matter model, and show how they are useful to get some constraints on the model parameters. Some useful formulae for the amplitudes of the scattering processes will be given in the Appendix.
\section{Minimal model with the singlet scalar}
The full Lagrangian considered in this paper simply consists of the SM Lagrangian~$\mathcal{L}_{\mathrm{SM}}$ and extra terms associated with the singlet scalar $S$,
\begin{equation}
\mathcal{L}=\mathcal{L}_{\mathrm{SM}} 
+\frac{1}{2}\partial_\mu S\, \partial^\mu S -\frac{1}{2} \mu_S^2\, S^2 + \frac{1}{4} \lambda_S\, S^4 + \frac{1}{2}\lambda_{HS} (H^\dagger H)\, S^2~,
\end{equation}
where  $H$ is the SM Higgs doublet and $\mathcal{L}_{\mathrm{SM}}$ contains the Higgs potential given as $-\mu^2 H^\dagger H+ \lambda_H (H^\dagger H)^2$.
Note that the singlet scalar S only couples to the SM Higgs $H$ among the SM particles and our results are irrespective of whether $S$ is a complex
or real singlet scalar.
Here, we consider two cases depending on whether the singlet scalar $S$ develops VEV or not.
As will be shown later, the implications on the unitarity condition depend on whether the VEV of $S$ is developed or not.
\subsubsection{Case for $<S>\,\neq 0$}

Let VEVs of the neutral components of $H$ and $S$ to be
$\langle H \rangle=\frac{1}{\sqrt{2}}v$ and 
$\langle S \rangle=\eta$, where  $v = (\sqrt{2} G_F )^{-1/2}$
and the value of $\eta$ is not determined from low energy experiments.
After two scalar fields $H$ and $S$ get VEVs,  they are written by
\begin{equation}
H=\left(
              \begin{array}{c}
               w^+ \\
               \frac{1}{\sqrt{2}}\Big(h+i\,z+v\Big) \\
              \end{array}
            \right)~~,~~~
S=\Big( s + \eta\Big)~,
\end{equation}
where the Goldstones $w^{+}$, $z$ are eaten by charged and neutral weak gauge bosons, $W$ and $Z$, in the SM, respectively.
Substituting these into the Lagrangian, we obtain mixing terms between two neutral fields $h$ and $s$ which are superpositions of two physical states $(h_1, h_2)$
given as follows:
\begin{equation}
\left(
  \begin{array}{c}
    h \\
    s \\
  \end{array}
\right)=
\left(
  \begin{array}{cc}
    \cos \alpha & -\sin \alpha \\
    \sin \alpha & \cos \alpha  \\
  \end{array}
\right)
\left(
  \begin{array}{c}
    h_1 \\
    h_2 \\
 \end{array}
\right)~,
\end{equation}
where the mixing angle $\alpha$ $(-\pi/2 \leq \alpha \leq \pi/2)$ is given by
\begin{equation}
\tan \alpha =\frac{-2 (\lambda_H c_\beta ^2 - \lambda_S\, s_\beta^2) \pm \sqrt{ 4(\lambda_H c_\beta^2 - \lambda_S\, s_\beta^2)^2 -\lambda_{HS}^2 c_\beta^2 s_\beta^2}}
{\lambda_{HS}\, c_\beta s_\beta}~,
\end{equation}
with $c_{\beta}\equiv \cos\beta=v/\sqrt{v^2+\eta^2}$, $s_{\beta}\equiv \sin\beta=\eta/\sqrt{v^2+\eta^2}$, and $\tan\,{\beta}=\eta/ v$.
For our convenience, we express three scalar quartic couplings  $\lambda_i$ in terms of the physical scalar masses, $m_{h_1}$ and $m_{h_2}$, and two mixing angles, $\alpha$ and $\beta$,
\begin{eqnarray}
\lambda_H &=& \frac{1}{4 c_\beta^2\,\xi^2}\left( m_{h_1}^2 c_\alpha^2 + m_{h_2}^2 s_\alpha^2 \right)~, \label{eq:lambda1}\\
\lambda_S &=& \frac{1}{4 s_\beta^2\,\xi^2}\left( m_{h_1}^2 s_\alpha^2 + m_{h_2}^2 c_\alpha^2 \right)~, \label{eq:lambda2}\\
\lambda_{HS} &=& \frac{s_{2\alpha}}{s_{2\beta}\,\xi^2}\left( m_{h_1}^2- m_{h_2}^2 \right)~, \label{eq:lambda3}
\end{eqnarray}
where $\xi^2=v^2 + \eta^2$.
We will assume that $h_1$ is always lighter than $h_2$, and denote their masses as $m_h$ and $m_s$, respectively.
Finally, the conditions derived from the fact that the potential should be bounded from below and all masses squared be positive are given by
\begin{equation}
\lambda_H > 0~,~~\lambda_S > 0~,~~ 4\,\lambda_H\, \lambda_S \geq \lambda_{HS}^2~,~~ 2\lambda_H\, c_\beta^2+2\lambda_S\, s_\beta^2 > 0~. \label{eq:VacuumStability}
\end{equation}
\subsubsection{Case for $<S>\,= 0$}
Imposing $\mathbb{Z}_2$ symmetry where only singlet $s$ is $\mathbb{Z}_2$-odd charged while other all fields are $\mathbb{Z}_2$-even charged, the singlet $s$ can not get a nontrivial VEV and can be regarded as a good candidate for dark matter.
As will be shown later, the size of the S-matrix is reduced in this case because
some scattering channels are forbidden by the $\mathbb{Z}_2$ symmetry.

It is important to notice that there are no bi-linear mixing terms ($\sim h s$) between $h$ and $s$ because the singlet $s$ does not develop the VEV.
In this case, the mass of singlet scalar $s$ is given by
\begin{equation}
m_s^2=\mu_s^2 + \lambda_{HS}\,\frac{v^2}{2}~.
\end{equation}
Contrary to the previous case ( $<S>\,\neq 0$ ), $m_s$ has nothing to do with $\lambda_H$.

The vacuum stability gives rise to the same conditions as in the previous case except for the third one in the Eq.\,(\ref{eq:VacuumStability}).
Requiring that the vacuum is located at the global minimum of the potential, we get the inequality given by
\begin{equation}
0< \mu_s^2 < \sqrt{\lambda_S \lambda_{HS}}~v^2 . \label{vac-st}
\end{equation}
\section{Unitarity of S-Matrix and numerical analysis}
Now let us consider various two-body scattering processes to derive the perturbative unitarity bound.
Before calculating the two-body scattering amplitudes, recall that the eigenvalues of the S-matrix does not depend on the choice of basis of the states.
So, for our convenience, we take weak eigenstates instead of mass eigenstates in the calculation simply because only scalar field $h$ couples to the gauge bosons.
Besides, since three external longitudinal gauge bosons can be replaced by corresponding Goldstone modes thanks to Goldstone-boson equivalence theorem, the amplitudes for two-body scattering processes we consider are equivalent to those with longitudinal gauge bosons up to terms of $O(M_W ^2 / s)$ which are negligible when $s \gg M_W^2$.

With the help of the partial wave decomposition, the scattering amplitude $\mathcal{M}$ is written by
\begin{equation}
\mathcal{M}(s,t,u)=16\pi \sum_{J=0}^{\infty}(2J+1)P_J(cos \theta)\, a_J(s), \label{scattering1}
\end{equation}
where $s, t, u$ are Mandelstam variables, $a_J(s)$ is the spin $J$ partial wave and $P_J$ are Legendre Polynomials.
The differential cross section is given by
\begin{equation}
\frac{d \sigma}{d \Omega}=\frac{1}{64\,\pi^2 s}|\mathcal{M}|^2~,
\end{equation}
and  by using the orthogonality of Legendre polynomial the cross section becomes
\begin{equation}
\sigma =\frac{16\pi}{s} \sum_{J=0}^{\infty}(2J+1)|a_J|^2~.
\end{equation}
 Applying the optical theorem that the cross section is proportional to the imaginary part of the amplitude in the forward direction, $\mathcal{M}(\theta=0)$, given by
\begin{equation}
\sigma = \frac{1}{s}\Im [\mathcal{M}(\theta=0)]~,
\end{equation}
we obtain the following unitarity constraint,
\begin{equation}
|a_J|^2 = \Im (a_J)~,~~\mathrm{for~~all}~~J~.
\end{equation}
It leads to the famous unitarity constraint of the partial wave amplitude $a_J$ with the identity $\Re(a_J)^2+\Im(a_J)^2=|a_J|^2$,
\begin{equation}
| a_J |^2 \leq \frac{1}{2} ~.
\end{equation}
The $J$th partial wave amplitude can be obtained by inverting Eq.\, (\ref{scattering1}),
\begin{equation}
a_J(s)=\frac{1}{32\pi}\int^1_{-1} dz P_J(z)\mathcal{M}(s,t,u)~,
\end{equation}
where the $z$ is the cosine of scattering angle.
To derive unitary bound, it is enough to focus on only $J=0$ s-wave amplitude $a_0(s)$ with vanishing external particle masses whose general form can be written by
\begin{equation}
a_0(s)=\frac{1}{16\pi}\Bigg[A
- B_h^2\left(\frac{1}{s-m_h^2}- \frac{\theta_t +\theta_s}{s} \ln (1 + \frac{s}{m_h^2})\right)
- B_s^2\left(\frac{1}{s-m_s^2}- \frac{\theta_t +\theta_s}{s} \ln (1 + \frac{s}{m_s^2})\right)
\Bigg]~,
\end{equation}
where A comes from the four point vertex, and the $B_h$($B_s$) is related with three point vertex with external $h$($s$) fields,
 and $\theta_t$, $\theta_s=0$ or 1 depending on the contributions of $t$ and $u$ channels in the process.
So, the upper bounds on the scalar masses can be derived from
\begin{equation}
|a_0| \leq \frac{1}{2}~.\label{eq:a0}
\end{equation}
\subsection{Case for $<S>\,\neq 0$}
\subsubsection{Limit of $s \gg m_{h}^2\,$, $m_s^2$}\label{sec:1}
The neutral states contributing to the scattering amplitudes are $|W^+W^- \rangle$, $|\frac{1}{\sqrt{2}} ZZ \rangle$, $|\frac{1}{\sqrt{2}} hh \rangle$, $|\frac{1}{\sqrt{2}} ss \rangle$, $|\frac{1}{\sqrt{2}} hs \rangle$, $|hZ \rangle$  with suitable normalization factor 1 or $\frac{1}{\sqrt{2}}$.\,\footnote{In fact, there also exist charged states whose contributions are simply presented by block diagonal elements of $T_0$ leading to an eigenvalue 1/2 which can not affect our results and discussion.  So we do not consider those contributions here.}
Their contributions to the scattering amplitude can be presented by $6 \times 6$ matrix form, and we denote it as $T_0$.
We note that the largest eigenvalue of the matrix $T_0$ gives rise to the strongest bounds on their masses and couplings.
It is obvious that the existence of the states such as $|\frac{1}{\sqrt{2}} ss \rangle$ and $|\frac{1}{\sqrt{2}} hs \rangle$ can make the eigenvalues of $T_0$ different from those in the SM.

For the $s \gg m_{h}^2\,$, $m_s^2$, the matrix $T_0$ in the basis ($|W^+W^- \rangle$, $|\frac{1}{\sqrt{2}} ZZ \rangle$, $|\frac{1}{\sqrt{2}} hh \rangle$, $|\frac{1}{\sqrt{2}} ss \rangle$, $|\frac{1}{\sqrt{2}} hs \rangle$, $|hZ \rangle$) takes the following form,
\begin{eqnarray}
T_0 \longrightarrow \left(-\frac{\lambda_H}{4\pi} \right)\cdot
\left(
  \begin{array}{cccccc}
    1 & \frac{1}{\sqrt{8}} & \frac{1}{\sqrt{8}} & 0 & 0 & 0 \\
    \frac{1}{\sqrt{8}} & \frac{3}{4} & \frac{1}{4} & 0 & 0 & 0  \\
    \frac{1}{\sqrt{8}} & \frac{1}{4} & \frac{3}{4} & \frac{3}{4}B & 0 & 0 \\
    0 & 0 & \frac{3}{4}B & \frac{3}{4}A & 0 & 0 \\
    0 & 0 & 0 & 0 & \frac{3}{4}B & 0 \\
    0 & 0 & 0 & 0 & 0 & \frac{1}{2}  \\
  \end{array}
\right)~,
\end{eqnarray}
where the $A$ and $B$ correspond to the scattering processes $ss \rightarrow ss$ and $hh \rightarrow ss$, respectively.
It is easy to check that Feynmann diagrams involving four vertex couplings can only survive in that limit at the tree level
because other scattering  channels are suppressed by the  factor $1/s$ in the propagators.
The parameters $A$ and $B$ are given in terms of the couplings by
\begin{equation}
A \equiv \frac{\lambda_S}{\lambda_H}~~,~~~ B \equiv \frac{1}{6}\frac{\lambda_{HS}}{\lambda_H}.
\end{equation}
Note that they are the ratios of the singlet relevant quartic couplings $\lambda_H $ and $\lambda_{HS}$ to the SM quartic coupling $\lambda_H$.
Taking $A$ and $B$ to be zero, the matrix form becomes equivalent to the $4\times 4$ matrix form of the SM, and we get the well-known perturbative unitarity bound called Lee-Quigg-Thacker bound~\cite{Lee:1977yc, Lee:1977eg}\label{eq:SMHB} on the Higgs mass in the SM,
\begin{equation}
M_H \leq \left( \frac{8 \sqrt{2}\,\pi}{3 G_{F}} \right)^{\frac{1}{2}} \equiv M_{LQT} \approx 1 \mathrm{TeV}~,
\end{equation}
where $|a_0| \leq 1$ has been applied and the highest eigenvalue $3/2$ has been taken from the original $4 \times 4$ matrix.
The eigenvalues of the matrix $T_0$ are composed of 4 eigenvalues derived from the $4 \times 4$ sub-matrix located at the left upper part of $T_0$ and two diagonal components of $T_0$, $\frac{3}{4}B$ and $\frac{1}{2}$. From the $4\times 4$ sub-matrix, we can get the characteristic polynomial
given by
\begin{equation}
\left[\Lambda-\frac{1}{2}\right]
\left[ \Lambda^3 - (A+2)\Lambda^2 + \left(2A-B^2+\frac{3}{4}\right)\Lambda - \left(\frac{3}{4}A -\frac{5}{4}B^2\right)\right]=0~.\label{eq:Char1}
\end{equation}
\begin{figure}
\begin{center}
\includegraphics[width=0.50\textwidth]{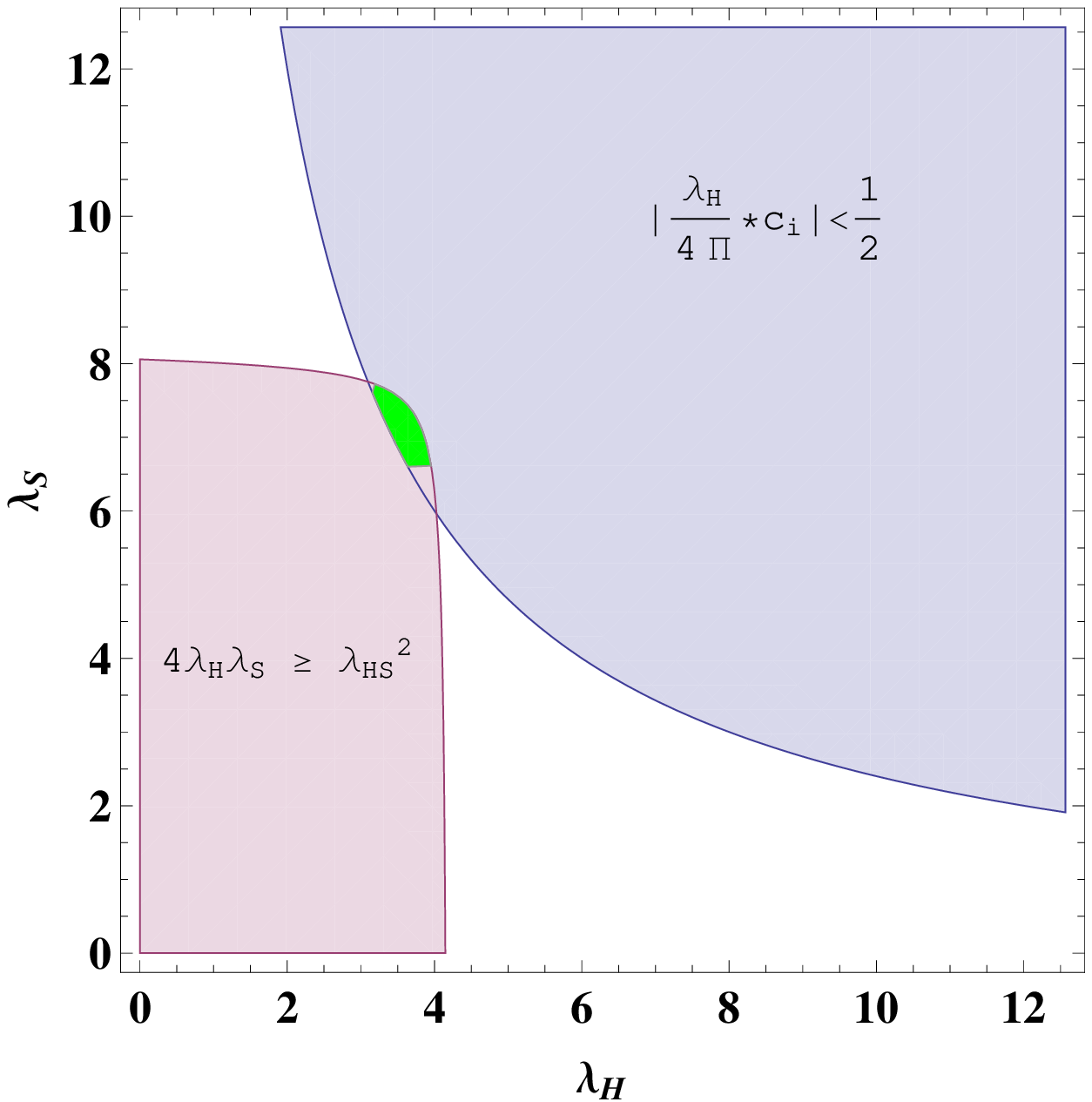}
\includegraphics[width=0.48\textwidth]{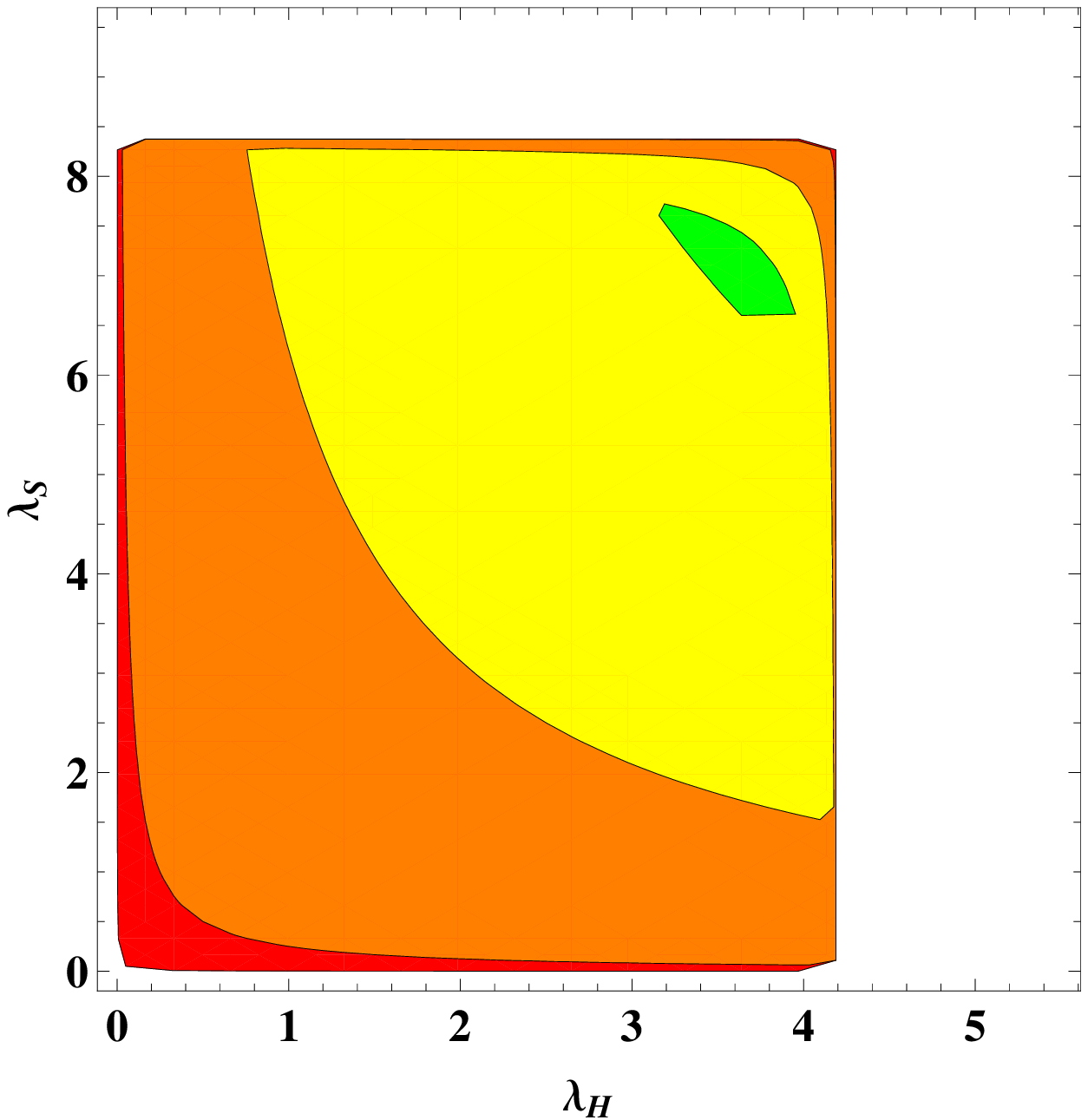}
\caption{Allowed regions by both vacuum stability and perturbative unitarity in the plain $(\lambda_H$ and $\lambda_S)$ for $\lambda_{HS}=$ 0 (red), 1 (orange), 5 (yellow) and 9.8 (greens).
The perturbative unitarity is imposed by taking the largest eigenvalues of $T_0$.
No allowed region exists for $\lambda_{HS} > 9.8$.}
\label{fig:Kappa}
\vspace{5pt}
\end{center}
\end{figure}
It is obvious that one solution of Eq.\,(\ref{eq:Char1}) is $1/2$, and the others are obtained by solving
the cubic equation with respect to $\Lambda$.
Since the cubic equation contains three unknown parameters, we first fix the value of $\lambda_{HS}$ and then numerically get the solutions
by varying the values of $\lambda_H$ and $\lambda_S$.
Once we obtain the eigenvalues of the matrix $T_0$ (denoted as $c_i$), we can derive the perturbative unitarity bound generally given by
\begin{eqnarray}
\left|\frac{\lambda_H}{4\pi}\cdot c_i \right| < \frac{1}{2}.\label{ineq:Uni}
\end{eqnarray}
Note that the above inequality with $c_i=1/2$ can naively be regarded as a perturbative condition on the coupling $\lambda_H$,  $\lambda_H \leq 4 \pi$. Thus, one can get stronger bound than the naive perturbative one as long as any eigenvalue of $T_0$ is larger than $1/2$.
With the help of Eqs.\,(\ref{eq:lambda1},\ref{eq:lambda2},\ref{eq:lambda3}), the bound on the coupling is translated into the bound on the mass given by
\begin{equation}
m_h^2 c_\alpha^2 + m_s^2 s_\alpha^2 < \frac{8\pi}{|c_i|}\xi^2 c_\beta^2
=\frac{8\pi v^2}{|c_i|}=\frac{4\sqrt{2}\,\pi}{G_F}\frac{1}{|c_i|}
=\frac{3}{2}\frac{1}{|c_i|}M_{LQT}^{\,2}~. \label{ineq:UniPhys}
\end{equation}

In the left panel of Fig.\,\ref{fig:Kappa}, we display how the regions of the parameter space in the plain
($\lambda_S, \lambda_H$) for a fixed value of $\lambda_{HS}$(=9.8) can be allowed by the vacuum stability and perturbative unitarity.
The purple and blue regions are allowed by the vacuum stability and
unitarity, respectively. Thus, the overlapped region is in consistent with both conditions.
The right panel of Fig.\,\ref{fig:Kappa} shows how the allowed region by both conditions varies with
different choice of $\lambda_{HS}$. The red, orange, yellow and green regions correspond to
$\lambda_{HS} = 0, 1, 5$ and $9.8$, respectively.
As $\lambda_{HS}$ increases, the allowed region gets narrower.
In our numerical analysis, we found that there is no allowed region in the plain ($\lambda_S, \lambda_H$)
for $\lambda_{HS}>9.8$.

\begin{figure}
\begin{center}
\includegraphics[width=0.45\textwidth]{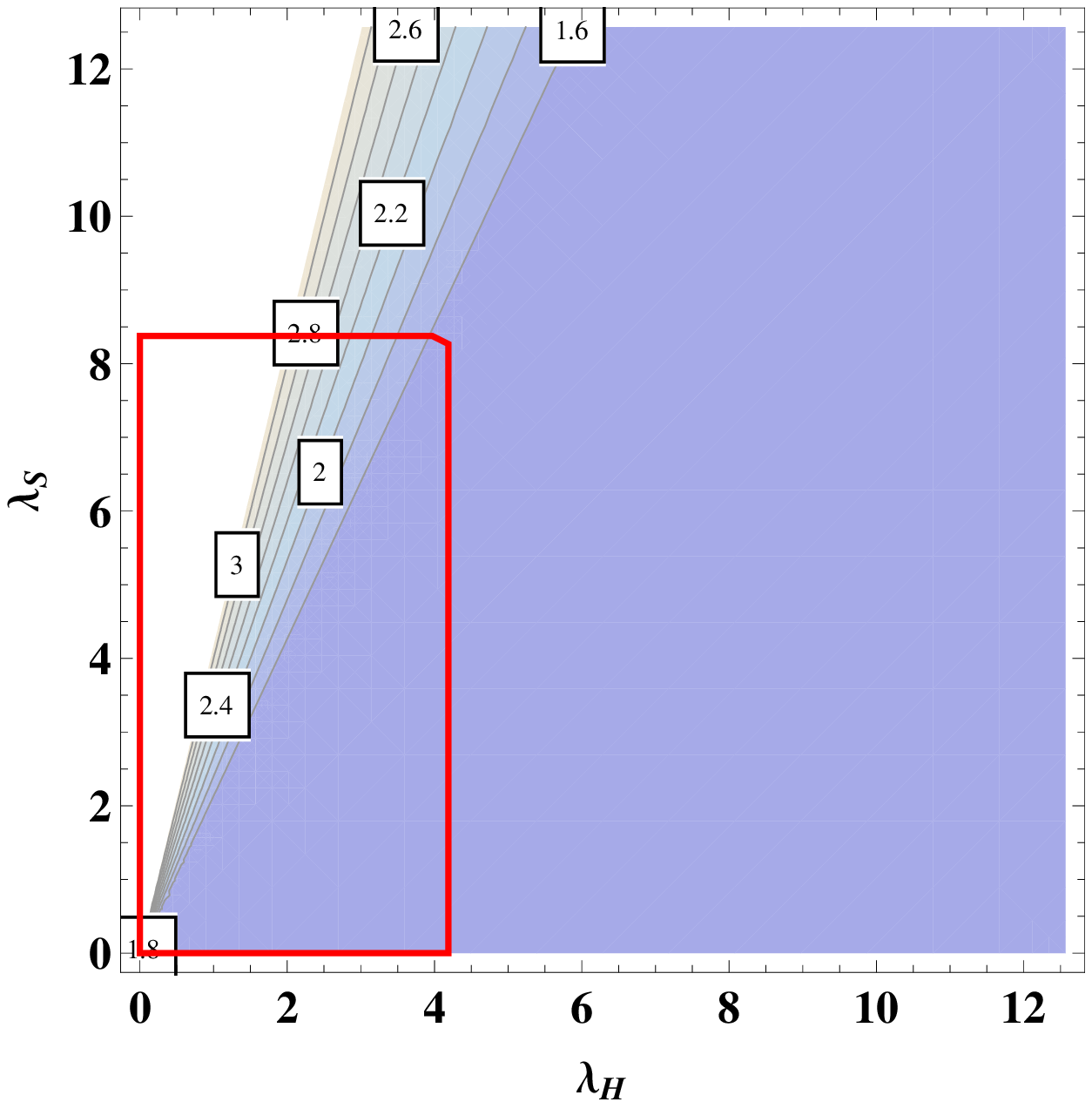}
\includegraphics[width=0.45\textwidth]{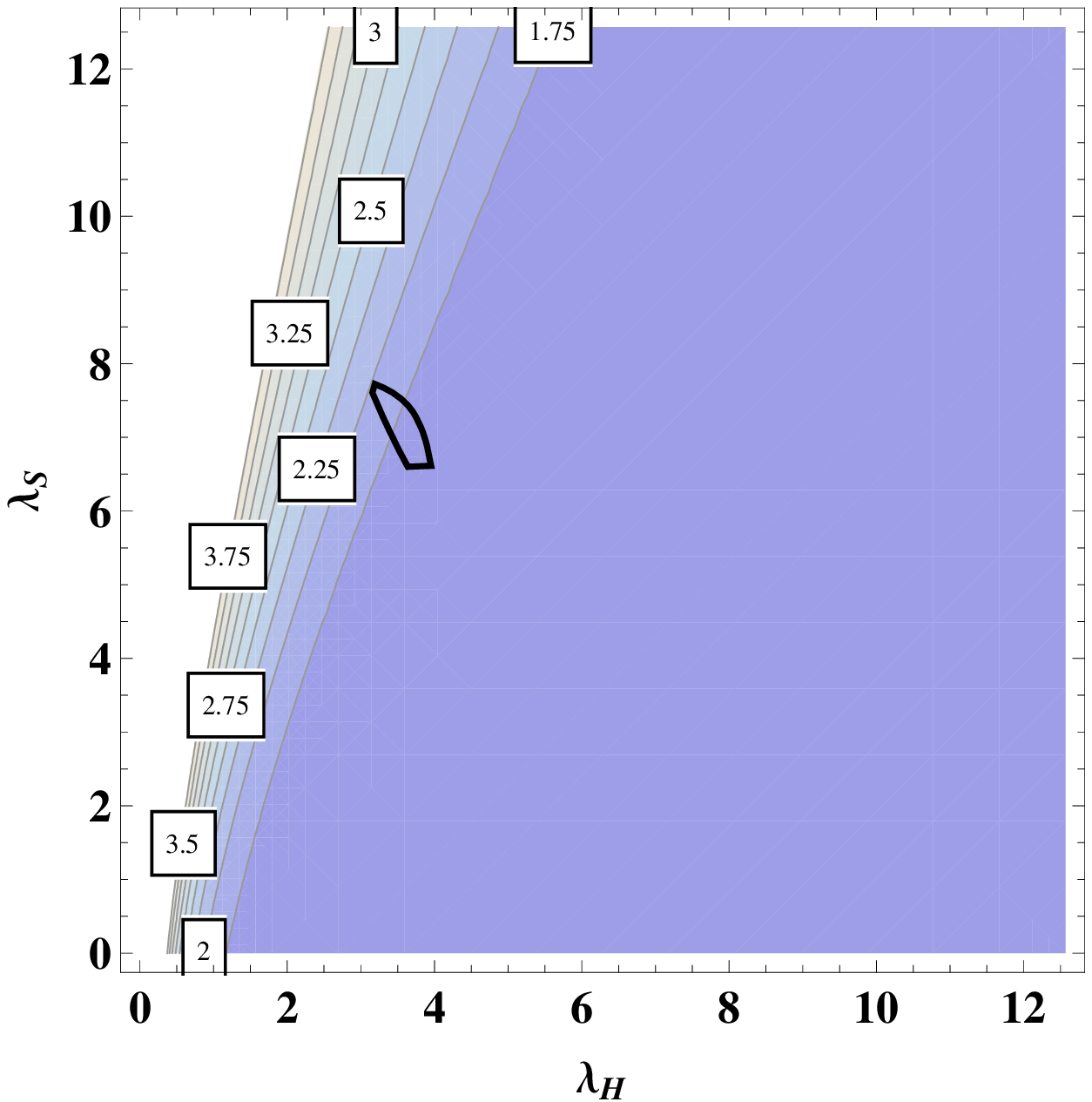}
\caption{Contour plots of the possible largest eigenvalues for Eq.(\ref{eq:Char1}) as a function of quartic couplings $\lambda_H$ and $\lambda_S$. Left(right) panel corresponds to $\lambda_{HS} \sim 0$($\lambda_{HS} = 9.8)$, and red and black lines denote the boundaries of allowed regions derived from the vacuum stability and perturbative unitarity bound, respectively.}
\label{fig:eigenvalues}
\vspace{5pt}
\end{center}
\end{figure}

Fig.\,\ref{fig:eigenvalues} shows how the eigenvalues of Eq.\,(\ref{eq:Char1}) are determined by varying both $\lambda_H$ and $\lambda_S$ for given value of $\lambda_{HS}$.
Notice that the contour plots displayed in Fig.\,\ref{fig:eigenvalues} correspond to the largest eigenvalues among three for fixed  $\lambda_{HS}$, whose numbers are presented in the rectangular boxes on each panels. The other two eigenvalues are not presented because they do not lead to stronger bounds.
The left (right) panels correspond to $\lambda_{HS}=0 (9.8)$.
We also show the allowed regions obtained by imposing the vacuum stability Eq.\,(\ref {eq:VacuumStability}) and perturbative unitarity condition Eq.\,(\ref{ineq:Uni}). The red and black curves represent the boundaries of the allowed regions.
As shown in Fig.\,\ref{fig:eigenvalues}, constraint by the vacuum stability is more severer than that by the perturbative unitarity in the case of
$\lambda_{HS}\sim 0$ , whereas vice verse in the case of $\lambda_{HS}=9.8$.
As mentioned before, the largest eigenvalues of the matrix $T_0$ can lead to the strongest
perturbative unitarity bound.
As can be seen from the panels in Fig.\,\ref{fig:eigenvalues}, the allowed largest eigenvalue is
reached to 3 (2) for $\lambda_{HS} \sim 0 (9.8)$, which in the end leads to much stronger unitarity bound,
compared with that in the SM where the largest eigenvalue is $3/2$.\,\footnote{
Since $T_0$ has at least an eigenvalue $1/2$, it is automatically satisfied with the perturbative condition of the quartic coupling given as $\lambda_H \leq 4 \pi$.}

\begin{figure}[h!t]
\includegraphics[width=0.61\textwidth]{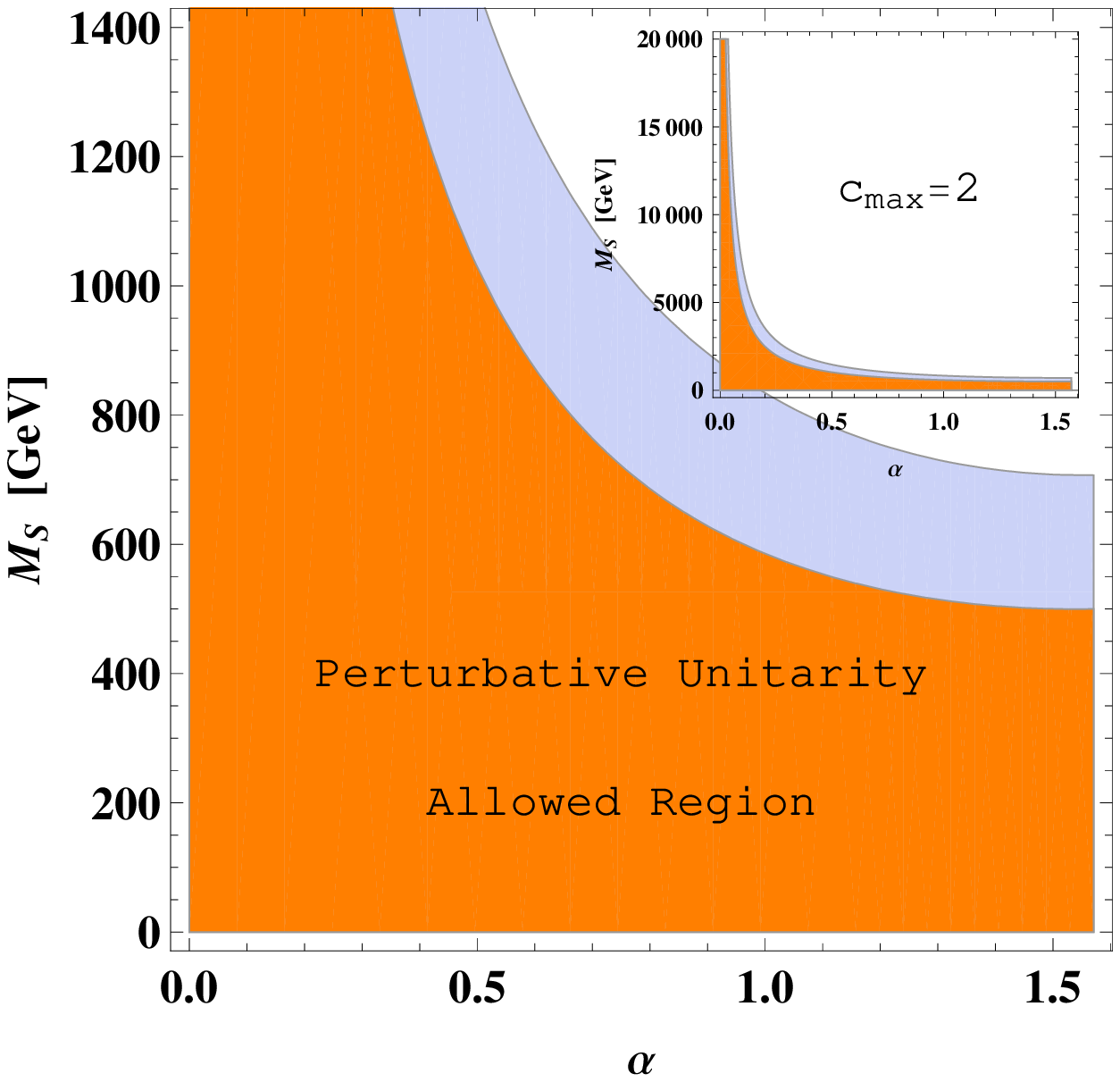}
\includegraphics[width=0.61\textwidth]{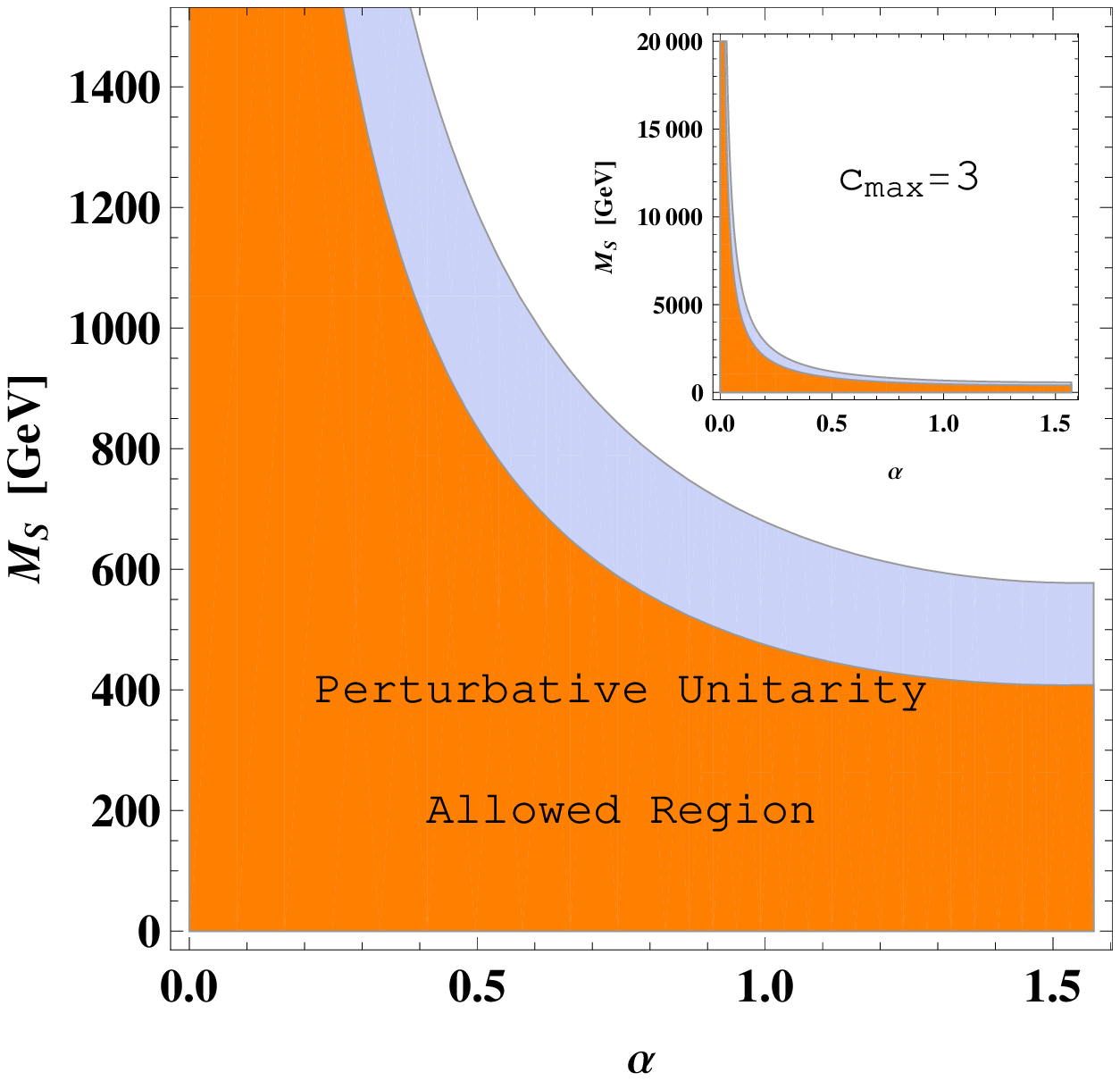}
\caption{Allowed regions of $m_s$ by the perturbative unitarity for $\mathrm{c}_{\mathrm{max}}=2$ (upper) and 3 (lower). In each panel the regions in grey and orange correspond to $|a_0| < 1$ and $|a_0|< 1/2$, respectively.
The insets in each panel show that the upper bounds on $m_s$ diverge when $\alpha \rightarrow 0$.} \label{fig:Scalar mass}
\vspace{5pt}
\end{figure}

As can be seen from the inequality\,(\ref{ineq:UniPhys}), the perturbative unitarity bound is translated as the mass bound
for the singlet scalar. Substituting the Higgs mass $m_h$ for the measured values of the boson mass at the LHC we can get upper bound on the mass of the singlet scalar.
Fig.\,\ref{fig:Scalar mass} shows the allowed region of $m_s$ by the perturbative unitarity along with the mixing angle $\alpha$ for the largest eigenvalues $\mathrm{c}_{\mathrm{max}}=2$ (upper panel) and $\mathrm{c}_{\mathrm{max}}=3$ (lower panel).
The grey (orange) region corresponds to $|a_0|<1$ ($|a_0|<\frac{1}{2}$). In each panel the grey region is introduced as a reference. It is obvious that the allowed region of $m_s$ for $|a_0| \leq 1/2$ is narrower than that for $|a_0| \leq 1$.
As can be seen from the insets of Fig.\,\ref{fig:Scalar mass}, for example, the bound on $m_s$ is around $5$ TeV for $\mathrm{c}_{\mathrm{max}}=2$, and $4$ TeV for $\mathrm{c}_{\mathrm{max}}=3$ in the case of $\alpha\sim 0.1$.
In the large mixing case ($\alpha \sim \pi/2$), we get very strong unitarity bounds on $m_s$. Numerically, they correspond to $1$ TeV $(\mathrm{c}_{\mathrm{max}}=1)$, $500$ GeV $(\mathrm{c}_{\mathrm{max}}=2)$, and $400$ GeV $(\mathrm{c}_{\mathrm{max}}=3)$, respectively.
Since the upper bound on $m_s$ diverges in the limit of mixing angle $\alpha \rightarrow 0$ as can be seen from two insets of Fig.\,\ref{fig:Scalar mass}, there is no bound on $m_s$ in the case that the light scalar field is perfectly the SM-like Higgs scalar.
Note that the unitarity bound on $m_s$ for $\alpha=\pi/2$ is 10 times larger than that for $\alpha=0.1$ because $m_s$ is multiplied by $s_\alpha^2$ in the (\ref{ineq:UniPhys}).

\subsubsection{Limit of $s \gg m_{h}^2$ and $s \sim 4 m_s^2 \sim (1~\mathrm{TeV})^2$}
The matrix $T_0$ in this limit takes the form
\begin{eqnarray}
T_0 \rightarrow \left(-\frac{\lambda_H}{4\pi} \right)\cdot
\left(
  \begin{array}{ccccccc}
    1 & \frac{1}{\sqrt{8}} & \frac{1}{\sqrt{8}} & \frac{1}{\sqrt{2}} C  & \frac{1}{\sqrt{2}} G & 0 \\
    \frac{1}{\sqrt{8}} & \frac{3}{4} & \frac{1}{4} & \frac{1}{2} C & \frac{1}{\sqrt{2}} G & 0  \\
    \frac{1}{\sqrt{8}} & \frac{1}{4} & \frac{3}{4} & \frac{3}{4} B & \frac{3}{4} F & 0 \\
    \frac{1}{\sqrt{2}} C & \frac{1}{2} C & \frac{3}{4} B & \frac{4}{3} A & \frac{3}{4} E & 0 \\
    \frac{1}{\sqrt{2}} G & \frac{1}{\sqrt{2}} G & \frac{3}{4} F  & \frac{3}{4}E & \frac{3}{4} D & 0 \\
    0 & 0 & 0 & 0 & 0 & \frac{1}{2}
  \end{array}
\right) \label{matrix2}
\end{eqnarray}
where $C,D,E,F$ and $G$ denote the factors of the amplitudes corresponding to the channels given by,
\begin{eqnarray}
&& C : a_0(W^+W^- \rightarrow ss) = a_0(ZZ \rightarrow ss)  ~,\\
&& D : a_0(hs \rightarrow hs) ~,\\
&& E : a_0(ss \rightarrow hs) ~,\\
&& F : a_0(hh \rightarrow hs) ~,\\
&& G : a_0(W^+W^- \rightarrow hs) = a_0(ZZ \rightarrow hs)~,
\end{eqnarray}
Their explicit amplitudes are presented in the appendix.
We note that the Feynman diagrams for the amplitudes associated with $C,E,F$ and $G$ contain propagators of the singlet scalar.
In the limit of large $s$, those amplitudes become negligible, so the matrix $T_0$ becomes the same form as in the case of $s \gg m_h^2, m_s^2$.

\begin{figure}
\begin{center}
\includegraphics[width=0.45\textwidth]{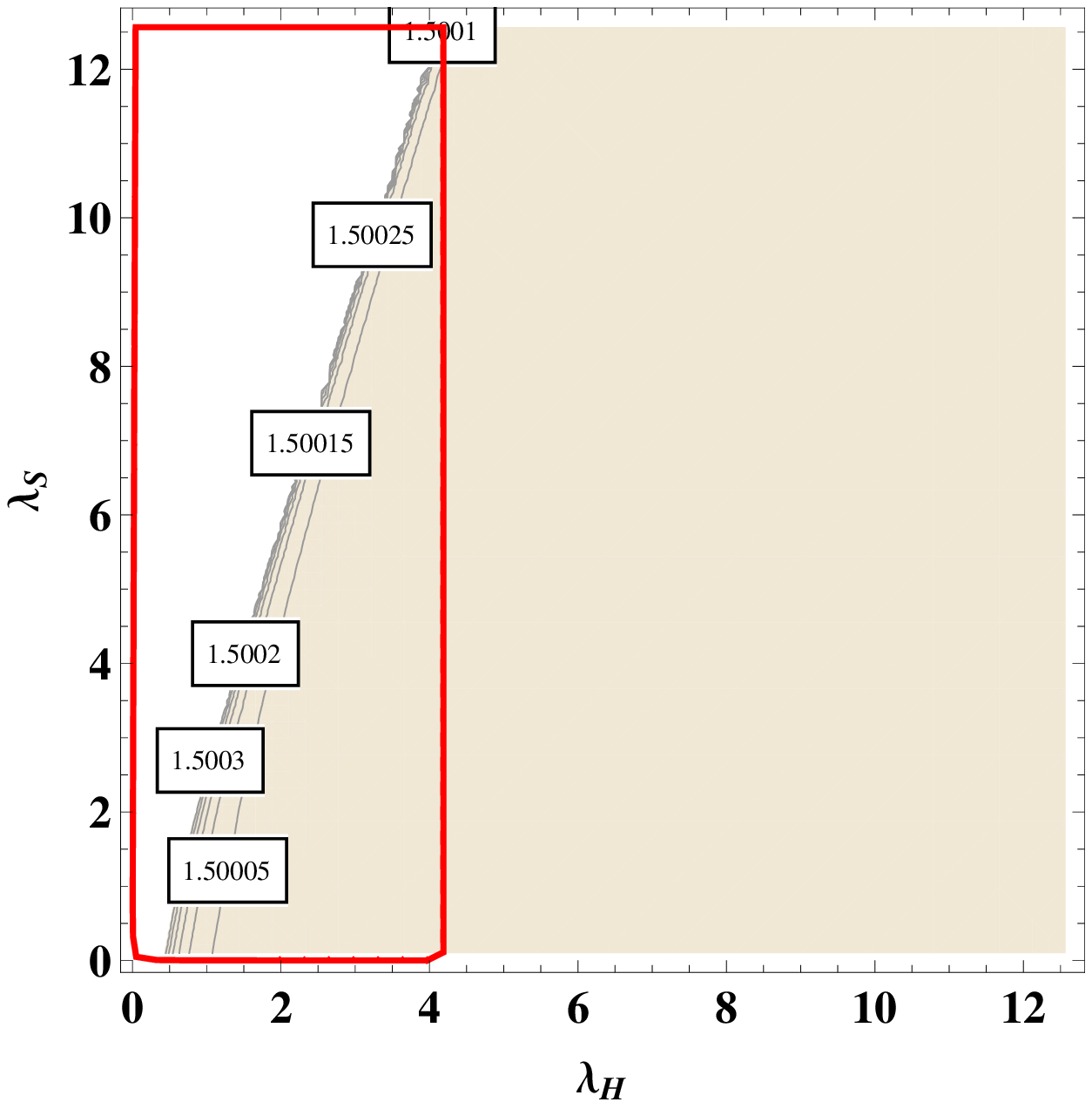}
\includegraphics[width=0.45\textwidth]{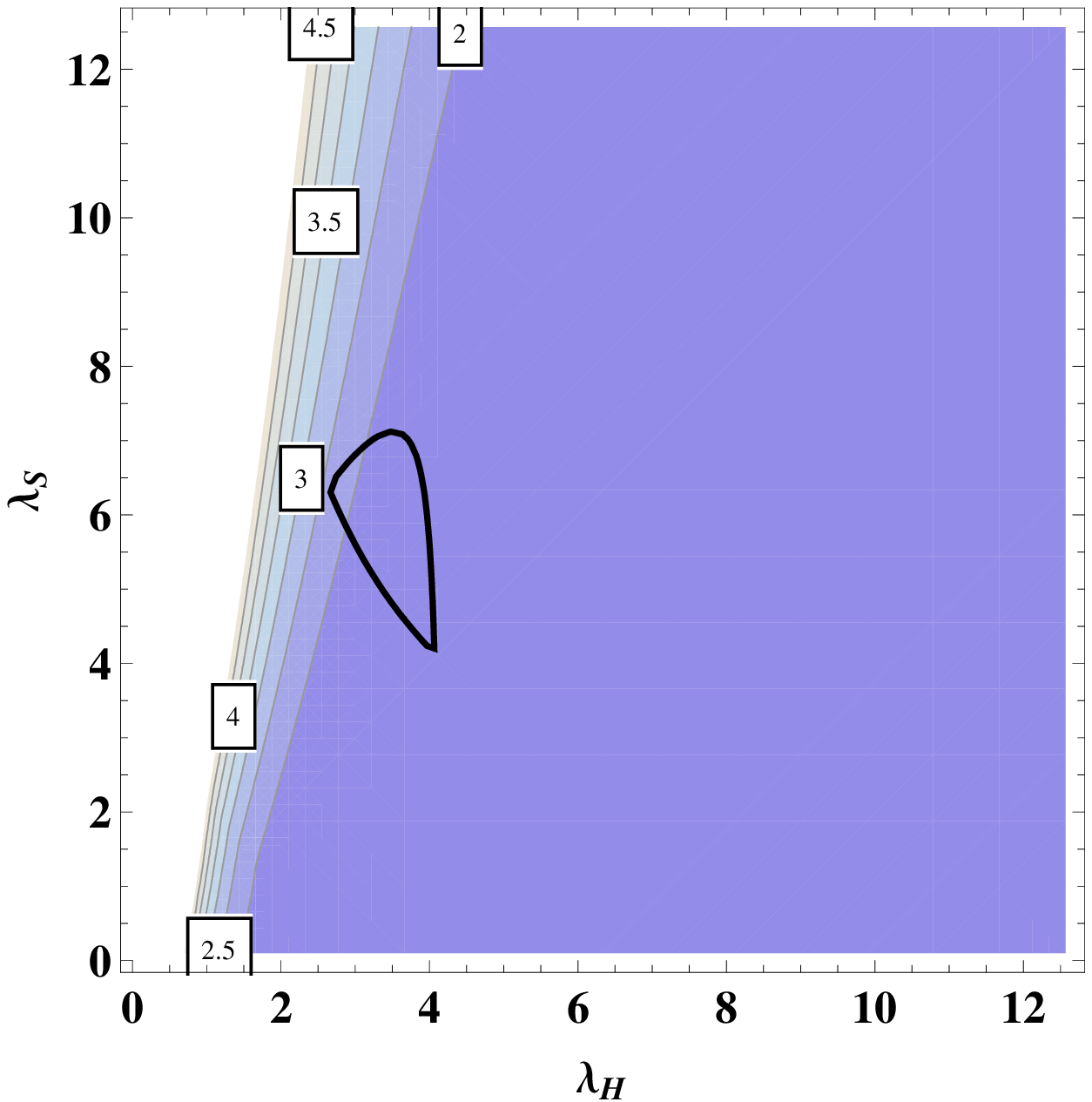}
\caption{The same as in Fig.\,\ref{fig:eigenvalues}, but left(right) panel corresponds to $\lambda_{HS} \sim 0.1$($\lambda_{HS} = 8.87)$
in the limit of $s \gg m_{h}^2$ and $s \sim 4 m_s^2 \sim (1~\mathrm{TeV})^2$.}
\label{fig:eigenvaluesGeneral}
\end{center}
\end{figure}

On top of an eigenvalue, $1/2$, directly taken from the diagonal component of $T_0$, we can obtain 5 eigenvalues
by solving characteristic equation for the non-diagonal $5\times 5$ sub-matrix located at the upper left
side of $T_0$.
Similar to the previous case, we display in Fig.\,\ref{fig:eigenvaluesGeneral} contour plots corresponding to the largest eigenvalues among five for both fixed $\lambda_{HS}$.
Here we choose  $s \sim 1~ \mathrm{TeV}$, and take $m_s(m_h)$ to be $450(126)$ GeV. Note that
Left(right) panel corresponds to $\lambda_{HS}=0.1(8.87)$,
where 8.87 is derived in the same way described in the previous subsection.
While most eigenvalues satisfying vacuum stability and perturbative unitarity are not larger than $3/2$ corresponding to the usual SM maximal eigenvalue, there exist several eigenvalues larger than $3/2$.
But, as can be seen from the right panel in Fig.\,\ref{fig:eigenvaluesGeneral}, the largest eigenvalue for $\lambda_{HS}=8.87$ is at best $2/5$.
Thus, the perturbative unitarity bound in this case appears to be weaker than that in the case $s \gg m_h^2, m_s^2$.
\subsection{Case for $<S>\,= 0$}
In this case, the matrix $T_0$ can be reduced to a simpler form because of the $\mathbb{Z}_2$-odd charge of singlet $s$.
It is worthwhile to notice that there is no {\it{s-h-h}} coupling because the singlet scalar $s$ can not develop the VEV, $\langle s \rangle = 0$, and
the odd parity of $s$ forbids the processes $W^+W^- \rightarrow hs$, $ZZ \rightarrow hs$, $hh \rightarrow hs$ and $hs \rightarrow ss$.
Thus, turning off the parameters $E,F$ and $G$ in the matrix form given by Eq.\,(\ref{matrix2}), we get
the matrix $T_0$ for this case as follows:
\begin{eqnarray}
T_0 \rightarrow \left(-\frac{\lambda_H}{4\pi} \right)\cdot
\left(
  \begin{array}{cccccc}
    1 & \frac{1}{\sqrt{8}} & \frac{1}{\sqrt{8}} & \frac{1}{\sqrt{2}} C  & 0 & 0 \\
    \frac{1}{\sqrt{8}} & \frac{3}{4} & \frac{1}{4} & \frac{1}{2} C & 0 & 0  \\
    \frac{1}{\sqrt{8}} & \frac{1}{4} & \frac{3}{4} & \frac{3}{4} B & 0 & 0 \\
    \frac{1}{\sqrt{2}} C & \frac{1}{2} C & \frac{3}{4} B & \frac{3}{4} A & 0 & 0 \\
    0 & 0 & 0 & 0 & \frac{3}{4} D & 0 \\
    0 & 0 & 0 & 0 & 0 & \frac{1}{2}
  \end{array}
\right)~.
\end{eqnarray}
\subsubsection{Limit of $s \gg m_{h}^2\,$, $m_s^2$}
In this limit, the elements of matrix $T_0$ proportional to $C$ vanish and only the tree-level four-point vertex contributions can remain. The parameter D becomes the same as $B$. Consequently, the form of the matrix $T_0$ becomes the same as that given in the previous subsec. \ref{sec:1}, so the largest eigenvalue of $T_0$ is $3$. However it just gives the upper bound on the Higgs mass because of no mixing between the Higgs and the singlet scalar.
Taking $\alpha=0$ and $\mathrm{c}_{\mathrm{max}}=3$,  we get the upper bound on the Higgs mass given as,
\begin{equation}
m_h \leq \frac{1}{\sqrt{2}}\,M_{LQT}~. \label{ineq:UniPhys2}
\end{equation}
Note that although there is no bound on the mass of the singlet $s$ in this case, there is still constraint on the coupling $\lambda_{HS}$ arisen from the same structure of $T_0$.
\subsubsection{Limit of $s \gg m_{h}^2$ and $s \sim 4 m_s^2$}
\begin{figure}
\begin{center}
\includegraphics[width=0.45\textwidth]{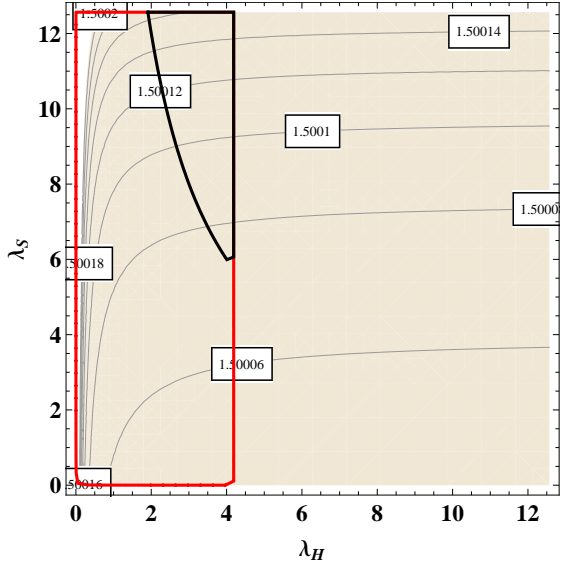}
\hspace{0.5cm}
\caption{Contour plots of the possible largest eigenvalues for Eq.(\ref{poly3}) as a function of quartic couplings $\lambda_H$ and $\lambda_S$.}
%
\label{fig:eigenvaluesZ2}
\vspace{5pt}
\end{center}
\end{figure}
The non-trivial characteristic polynomial for the upper $4 \times 4$ block of the matrix $T_0$ which has a trivial eigenvalue $1/2$ is given by
\begin{equation}
\Bigg( 64\Lambda^3 + (-128 - 48 A) \Lambda^2 + (48 + 96 A - 36 B^2 - 48 C^2)\Lambda -36 A + 45 B^2 - 36 B C + 36 C^2 \Bigg)=0~. \label{poly3}
\end{equation}
In Fig.\,\ref{fig:eigenvaluesZ2}, we display contour plots representing the largest eigenvalues obtained by numerically solving Eq.\,(\ref{poly3}).
Because of no mixing between the Higgs and singlet scalars, the coupling $\lambda_{HS}$ does not affect the determination of the eigenvalues at all, contrary to the previous cases.
Here we take $s ~\sim 1~\mathrm{TeV}^2$, $m_h=126$ GeV and $m_s=450$ GeV as in the previous subsection.
 We see from Fig.\,\ref{fig:eigenvaluesZ2} that the largest eigenvalue is determined to be around $3/2$ irrespective of the value of $\lambda_{HS}$.
Note that non-negligible matrix elements can lower the largest eigenvalue compared with the one in the limit of $s \gg m_h^2, m_s^2$.
The upper bound on $m_h$ corresponding to the largest eigenvalue $3/2$ is given by
\begin{equation}
m_h \leq \sqrt{2}\,M_{LQT}~. \label{ineq:UniPhys3}~
\end{equation}
\section{Implications and Conclusion}
Requiring perturbative unitarity of the S-matrix in the SM extended to contain a singlet scalar, we could get some bounds on the scalar masses.
In particular, we derived the upper bound on the singlet scalar mass by taking the Higgs mass to be 126 GeV
measured by the LHC. In Table \ref{tab:table1}, we summarize  the upper bounds on the scalar masses along with the limits of the center-of-mass energy $s$ (first column), the largest eigenvalues of $T_0$ (second),
the discrete symmetry of model (third), mixing angle $(\alpha)$ (fifth) and the coupling $\lambda_{HS}$ (sixth).
\begin{table}
\caption{Upper bounds on the scalar masses along with the limits of the center-of-mass energy $s$ (first column), the largest eigenvalues of $T_0$ (second),
the discrete symmetry of model (third), mixing angle $(\alpha)$ (fifth) and the coupling $\lambda_{HS}$ (sixth).} \label{tab:table1}
\begin{center}
\begin{tabular}{|c|c|c|c|c|c|}
               \hline
               Condition & \,$\mathrm{c}_\mathrm{max}$\,& \,Sym. & Mass bound & \,Mixing angle $(\alpha)$ & $\lambda_{HS}$ \\
               \hline \hline
               $s \gg m_h^2~,~ m_s^2$ & 3 & $\times$ &  $m_s \leq 4.5 ~\mathrm{TeV}(400 ~\mathrm{GeV})$ & $\alpha \sim 0.1(1.5)$ & 0 $\sim$ 9.8\\
               \hline
               $s \gg m_h^2~,~ s \sim 4m_s^2 \sim 1~\mathrm{TeV}$~ & 5/2 & $\times$ &  $m_s \leq 6 ~\mathrm{TeV}(550 ~\mathrm{GeV})$ & $\alpha \sim 0.1(1.5)$ & ~0 $\sim$ 8.87~\\
               \hline
               $s \gg m_h^2~,~ m_s^2$ & 3 &  $\mathbb{Z}_2$ & $m_h \leq (1/ \sqrt{2})\,M_{LQT} \approx ~707\,\mathrm{GeV}$ & $\times$ & 0 $\sim$ 9.8\\
               \hline
               $s \gg m_h^2~,~ s \sim 4m_s^2 \sim 1~\mathrm{TeV}$  & 3/2 & $\mathbb{Z}_2$ &  $m_h \leq \sqrt{2}\,M_{LQT} \approx ~1404\,\mathrm{GeV}$ & $\times$ & ~0 $\sim$ 8.87~\\
               \hline
\end{tabular}
\end{center}
\end{table}

Based on the upper bounds on the scalar masses we derived, let us discuss the implications of
those bounds on two interesting scenarios in which scalar fields play a crucial role in solving
problems of inflation and dark matter.
\subsection{Unitarized Higgs inflation}
Recently, it has been proposed that cosmic inflation can be driven by the SM Higgs with a large non-minimal coupling to Ricci scalar~\cite{Bezrukov:2007ep}, namely $\xi H^\dagger H R$ and $\xi\, \sim \, 10^4$. But soon it was pointed out that the original Higgs inflation model can be afflicted with the unitarity problem due to non-minimal Higgs couplings~\cite{Burgess:2010zq}. 
To resolve the unitarity problem while maintaining perturbativity up to the cut-off $(\Lambda)$ scale  of the model,
an additional gauge-singlet scalar is introduced~\cite{Giudice:2010ka} or appropriate counter terms are taken into account~\cite{EliasMiro:2012ay}.

Here we mainly concentrate on the unitarized (explicitly, the linear $\sigma$ model type) Higgs inflation model that has an additional singlet scalar $s$.
In this scenario, a state composed of both the Higgs and the singlet scalar plays the role of inflaton~\cite{Lebedev:2011aq}.
Therefore, this scenario requires nontrivial VEV of the singlet scalar so as to generate a mixing between
the singlet and Higgs scalars.
The relevant Lagrangian of the model in the Jordan frame is given by
\begin{equation}
\mathcal{L}_{\,\mathrm{Jordan}}/ \sqrt{-g}=-\frac{1}{2}M_{Pl}^2\, R -\frac{1}{2}\xi_{h}\,h^2 R -\frac{1}{2}\xi_s\, s^2 R
 + \frac{1}{2}(\partial_{\mu} h)^2 + \frac{1}{2}(\partial_{\mu} s)^2 - V(h,s)
\end{equation}
where $\xi_{h, \, s} > 0 $ are dimensionless parameters that can control the inflation in the early universe at the large field value.
The scalar potential $V(h,s)$ in this scenario has the following form,
\begin{equation}
V=\frac{1}{4}\lambda_h \, h^4 + \frac{1}{4}\lambda_s \, s^4 + \frac{1}{4}\lambda_{hs} h^2 s^2 + \frac{1}{2}m_h^2 h^2 + \frac{1}{2}m_s^2 s^2.
\end{equation}
Note that the potential is exactly the same as the one given by Eq.\,(1) \footnote{For a review on the unitarized Higgs inflation model, see the Ref. ~\cite{Lebedev:2011aq}.}.

\begin{figure}
\begin{center}
\includegraphics[width=0.7\textwidth]{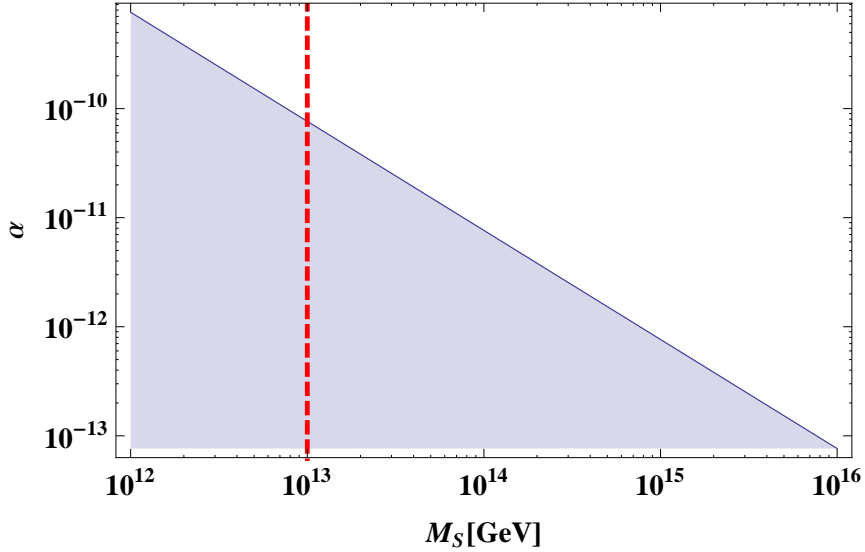}
\caption{Allowed (blue shaded) region of $\alpha$ {\it vs.} $m_s$ for the eigenvalue $c=5/2$.
The red dashed line corresponds to the upper bound on $m_s$ coming from  the COBE constraint~\cite{EliasMiro:2012ay}.}
\label{fig:Alpha}
\vspace{5pt}
\end{center}
\end{figure}

Using Eq.\,(\ref{ineq:UniPhys}) in this scenario, we can easily obtain an inequality for the mixing angle $\alpha$ given as
\begin{equation}
\alpha < \sin^{-1} \Bigg[ \Bigg(\frac{3 M_{LQT}^2}{2\, c_i} - m_h^2 \Bigg) / \Big( {m_s^2 - m_h^2} \Big) \Bigg]~\label{ineq:angle},
\end{equation}
In Fig.\,\ref{fig:Alpha}, we plot the upper bound on the mixing angle $\alpha$ as a function of the mass of the singlet scalar $S$ in the limit of $m_s \gg m_h$ after taking the Higgs mass to be 126 GeV.
Interestingly, in Fig.\,\ref{fig:Alpha}, we can easily see that the mixing angle $\alpha$ should be very small,
\begin{equation}
\alpha \leq [10^{-9}, 10^{-13}\,]~~~ \mathrm{on}~~~ M_s \in [10^{12}, 10^{16}\,]~.
\end{equation}
Imposing the COBE result for normalization of the power spectrum ~\cite{Lyth:1998xn} on the parameters, we can get the relation,
\begin{equation}
\frac{\sqrt{\lambda_s}}{\xi_s}=2 \times 10^{-5} \sqrt{\frac{\lambda_h}{\lambda_h-\lambda_{hs}^2 / \lambda_s}}~,
\end{equation}
which is translated into the mass relation for the singlet scalar $S$ given by \cite{EliasMiro:2012ay}
\begin{equation}
M_s^2 \simeq \lambda_s\, \frac{M_{Pl}^2}{3\,\xi_s^2},
\end{equation}
where $M_{Pl}$ is the Planck Mass in the given model. From the above COBE constraint, we get $M_s \approx 10^{13}$ GeV and
it is represented in Fig.\,\ref{fig:Alpha} by a dashed red line.
We see from Fig.\,\ref{fig:Alpha} that the COBE constraint leads to the upper bound on the mixing angle, $\alpha \leq 10^{-10}$.
From our numerical analysis, we found that the change of eigenvalue from $\frac{5}{2}$ to $3$ does not affect the allowed range of $\alpha$.
In fact, it is obvious that such a tiny value of $\alpha$ comes from a big mass hierarchy between the Higgs and the singlet scalars in this scenario.

\subsection{TeV scale singlet dark matter}
Regarding the singlet scalar as a TeV scale dark matter candidate (DM)~\cite{Ponton:2008zv}, let us discuss
how the perturbative unitarity condition can constrain the model parameter by combining it with the relic density of DM.
The annihilation cross section of the singlet scalar DM into two Higgs bosons in the limit of $m_s \gg m_h$ is
simply given by
\begin{equation}
\langle \sigma_{ss\rightarrow hh}~v \rangle \approx \frac{\lambda_{HS}^2}{16\pi m_s^2}~,
\end{equation}
where the $v$ is the relative velocity of the annihilation particles, and the bracket denotes the thermal average.
Note that the above annihilation channel is dominant over the other annihilation channels such as $ss\rightarrow ww/zz$ in the case of $m_s \sim 1$ TeV.
We see that the relic density of the singlet scalar DM depends on the coupling $\lambda_{HS}$ and its mass $m_s$.
Combining the unitarity constraint on the coupling $\lambda_{HS}$ with the measurement of the relic density, we can derive some bound on the mass of the singlet scalar.
From the 9\,-year WMAP result for the cold dark matter density given by  $\Omega_{DM}\,h^2 = 0.1138 \pm 0.0045$~\cite{Bennett:2012fp},
we obtain the following relation
\begin{equation}
\Omega_{DM}\,h^2 \approx \frac{1.04 \times 10^5 \mathrm{GeV^{-1}}}{M_P}\frac{x_F}{\sqrt{g_*}}\frac{1}{\langle \sigma_{ss\rightarrow hh}~v \rangle}~,
\end{equation}
where $M_P$ is the Planck mass ($\approx 1.22 \times 10^{19}$ GeV), $x_F=m_S/T_F$ with the freeze-out temperature $T_F$, and the $g_*$ the effective number of relativistic degrees of freedom at freeze-out. The suitable values of $x_F$ and $g_*$ are about 25 and 90, respectively.\,\footnote{see Ref. \cite{Ponton:2008zv} for the details on the singlet scalar dark matter model and the constants required in the calculation of the DM relic density.}
\begin{figure}
\begin{center}
\includegraphics[width=0.50\textwidth]{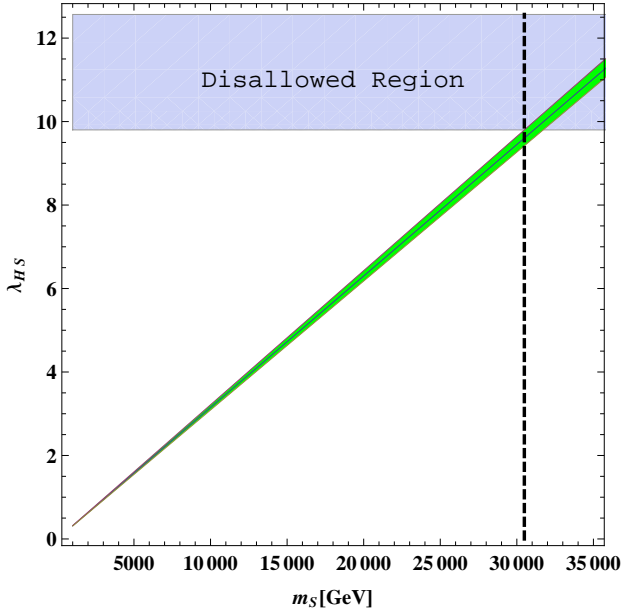}
\caption{Allowed region of the parameter space in the plain ($m_s, \lambda_H$) from the observed DM abundance at $1 \sigma$ C.L.
The blue region is disallowed by the unitarity condition, and the black dashed line represents the upper bound on $m_s \leq ~ 30.490~\mathrm{TeV}$.}
\label{fig:Mupper}
\vspace{5pt}
\end{center}
\end{figure}
Fig.\,\ref{fig:Mupper} shows how the upper bound on the singlet scalar mass as a DM candidate can be determined by imposing the perturbative unitarity constraint on $\lambda_{HS}$ to the prediction of the relic density of DM. In Fig.\,\ref{fig:Mupper}, the green band represents the allowed region of the parameter space in the plain
($m_s, \lambda_{HS}$) from the observed DM abundance at $1 \sigma$ C.L., and the blue region is the disallowed region coming from the unitarity constraint.
We display the black dashed line representing the upper bound on $m_s$ determined from the combination of the observed DM abundance and unitarity bound.
\begin{equation}
m_s \leq ~ 30.490~\mathrm{TeV}\, . \footnote{In Ref.\cite{Ponton:2008zv}, the authors have studied TeV scale singlet scalar dark matter by restricting the cutoff scale of the model to be a few TeV.
But, our analysis shows that there exists a valid perturbative regime up to 30 TeV.}
\end{equation}


In conclusion, we have studied the implication of the perturbative unitarity in the SM extended to include the singlet scalar particle.
Taking into account full contributions to the scattering amplitudes, we have derived unitarity conditions  on the S-matrix which can be translated into bounds on the masses of the scalar fields.
In the case that the singlet scalar field develops vacuum expectation value (VEV), we could get
the upper bound on the singlet scalar mass varying with the mixing angle between the singlet and Higgs scalars.
While the bound becomes divergent in the decoupling limit ($\alpha \rightarrow 0$),  the bound becomes very strong, $m_s \lesssim 400$ GeV, as the mixing angle $\alpha$ reaches maximal.
On the other hand, the mass of the Higgs scalar can be constrained
by the unitarity condition in the case that the VEV of the singlet scalar is not generated.
We found that the unitarity bound on the Higgs mass is modified and can appear to be severer in the presence of the singlet scalar field.
We have shown how the unitarity condition can constrain the unitarised Higgs inflation, and
found that a tiny mixing angle $\alpha \sim 10^{-10}$ is required for the singlet scalar with around $10^{13}$ GeV in the model.
The singlet scalar mass is not constrained by the unitarity itself when we impose $Z_2$ symmetry in the model because of no mixing with the Higgs scalar.
But, regarding the singlet scalar field as a cold dark matter candidate, we have derived upper bound on the singlet scalar mass, $m_s \lesssim ~ 30~\mathrm{TeV}$, by combining the observed relic abundance with the unitarity.

%

\newpage
\appendix{\bf{$\langle$Appendix: amplitudes of scattering processes$\rangle$}}\\
We here present the explicit S-wave partial amplitudes to calculate the perturbative unitarity bound
in the convention of ref.~\cite{Cynolter:2004cq}. First the $\mathbb{Z}_2$-charge conserving transitions are:
\begin{eqnarray}
&&\bullet~ a_0(hh \rightarrow hh)
= -\frac{\lambda_H}{4 \pi}\,\kappa_{hh \rightarrow hh} \left(\frac{3}{2}\right)
\Bigg[ 1 + \frac{3 m_h^2}{s-m_h^2} - \frac{6 m_h^2}{s - 4m_h^2}\ln \left( \frac{s}{m_h^2} -3 \right) \nonumber \\
&&+ \delta^2 \tan^2 \beta \left( \frac{3 m_h^2}{s-m_s^2}-\frac{6m_h^2}{s-4m_h^2}\ln \left(\frac{s-4m_h^2}{m_s^2}+1\right) \right) \Bigg]~,\\
&&\bullet~ a_0(ss \rightarrow ss)
= -\frac{\lambda_H}{4 \pi}\, \kappa_{ss \rightarrow ss} \left(\frac{3}{2}\right)
\Bigg[ \tilde{\delta} + \delta^2\left( \frac{3m_h^2}{s-m_h^2}-\frac{6m_h^2}{s-4m_s^2}\ln \left(\frac{s-4m_s^2}{m_h^2}+1 \right) \right) \nonumber \\
&& + \tilde{\delta}^2 \tan^2 \beta \left( \frac{3 m_h^2}{s-m_s^2}-\frac{6m_h^2}{s-4m_s^2}\ln \left(\frac{s}{m_s^2}-3 \right) \right) \Bigg]~, \\
&& \bullet~ a_0(ss \rightarrow hh)
= \frac{\lambda_H}{4 \pi}\, \kappa_{ss \rightarrow hh} \left(\frac{3}{2}\right)
\Bigg[\delta + \delta \frac{3 m_h^2}{s-m_h^2} -\delta^2 \frac{6 m_h^2}{\sqrt{s - 4m_h^2}\sqrt{s -4m_h^2}} \nonumber \\
&&\times \ln \left( 1+ \frac{2 \sqrt{s - 4m_h^2}\sqrt{s -4m_h^2}}{s- \sqrt{s - 4m_h^2}\sqrt{s -4m_h^2} -2 m_h^2} \right)
+\delta \tilde{\delta} \tan \beta \frac{3m_h^2}{s-m_s^2} \nonumber \\
&&-\delta^2 \tan^2 \beta \frac{6m_h^2}{\sqrt{s - 4m_h^2}\sqrt{s -4m_h^2}}
\ln \left( 1 + \frac{2 \sqrt{s- 4m_s^2} \sqrt{s- 4m_h^2} } { s - \sqrt{s- 4m_s^2} \sqrt{s- 4m_h^2} -2 m_h^2} \right) \Bigg] ~, \\
&& \bullet~ a_0(hs \rightarrow hs)
= -\frac{\lambda_H}{4 \pi}\, \kappa_{hs \rightarrow hs} \left(\frac{3}{2}\right)
\Bigg[ \delta + \delta^2\frac{3m_h^2}{s-m_s^2} - \delta \frac{3m_h^2 s}{A}
\ln \left(\frac{s^2-s\,(2m_s^2+m_h^2)+(m_h^2-m_s^2)^2}{s\, m_h^2}\right) \nonumber \\
&&-\delta^2 \frac{3m_h^2 s}{A} \ln \left( \frac{s^2-s(2m_s^2+m_h^2)}{s m_s^2 - (m_h^2-m_s^2)^2} \right)
+\delta^2 \frac{3m_h^2}{s-m_h^2} \nonumber \\
&& -\delta\tilde{\delta}\tan^2\beta \frac{3m_h^2\,s}{A} \ln \left( \frac{s^2-s(2m_h^2+m_s^2)+(m_h^2 - m_s^2)^2}{s\,m_s^2} \right)
-\delta^2 \frac{3m_h^2\,s}{A} \ln \left( \frac{s^2-s(m_h^2+2 m_s^2)}{s m_h^2 - (m_h^2-m_s^2)^2} \right) \Bigg] ~, \\
&& \bullet~ a_0(hh \rightarrow W^+_L W^-_L) = a_0(hh \rightarrow Z_L Z_L)
= -\frac{\lambda_H}{4 \pi}\, \kappa_{hh \rightarrow ww/zz} \left(\frac{1}{2}\right) \Bigg[ 1 + \frac{3 m_h^2}{s-m_h^2} \nonumber  \\
&&-\frac{4 m_h^2}{[s(s-4 m_h^2)]^{1/2}} \ln \left( \frac{s -2m_h^2-[s(s-4 m_h^2)]^{1/2}}{2m_h^2} \right) \Bigg] ~, \\
&& \bullet~ a_0(ss \rightarrow W^+_L W^-_L) = a_0(ss \rightarrow Z_L Z_L)
=- \frac{\lambda_H}{4 \pi}\, \kappa_{ss \rightarrow ww/zz} \left(\frac{1}{2}\right)\left(\delta\frac{3 m_h^2}{s-m_h^2}\right)~,\\
&& \bullet~ a_0(h Z_L \rightarrow h Z_L)
= -\frac{\lambda_H}{4 \pi} \kappa_{hz \rightarrow hz} \left(\frac{1}{2}\right)
\Bigg[ 1 + \frac{m_h^2}{s} - \frac{3m_h^2\,s}{(s-m_h^2)^2} \ln \left(1 + \frac{ (s - m_h^2)^2}{s m_h^2}\right) \nonumber \\
&& - \frac{s\,m_h^2}{(s-m_h^2)^2} \ln \left( \frac{ s (2m_h^2 -s )}{m_h^4}\right) \Bigg] \\
&& \bullet~ a_0(h Z_L \rightarrow hh)=0~,~~~~~~\bullet\, a_0(h Z_L \rightarrow Z_L Z_L)=0~,~~~~~~\bullet\, a_0(h Z_L \rightarrow W^+_L W^-_L)=0~.
\end{eqnarray}
where $A=s^2 -2s (m_h^2+m_s^2) + (m_h^2 - m_s^2)^2$ and the kinematic factor $\kappa_{AB \rightarrow CD}$ is defined by
\begin{equation}
\kappa_{AB \rightarrow CD} \equiv
\left( 1 - \frac{(m_A - m_B)^2}{s}\right)^{\frac{1}{4}}\left( 1 - \frac{(m_A + m_B)^2}{s}\right)^{\frac{1}{4}}
\cdot
\left( 1 - \frac{(m_C - m_D)^2}{s}\right)^{\frac{1}{4}}\left( 1 - \frac{(m_C + m_D)^2}{s}\right)^{\frac{1}{4}}~.\nonumber
\end{equation}

Secondly, the $\mathbb{Z}_2$-charge violating processes :
\begin{eqnarray}
&& \bullet~ a_0(hs \rightarrow W^+_L W^-_L) = a_0(hs \rightarrow ZZ)
=- \frac{\lambda_H}{4 \pi}\, \kappa_{hs \rightarrow ww/zz} \left(\frac{1}{2}\right)\left(\delta \tan \beta \frac{3 m_h^2}{s-m_h^2}\right)\nonumber \\
&& \bullet~ a_0(hh \rightarrow hs) = - \frac{\lambda_H}{4\pi}\, \kappa_{hh \rightarrow hs}\left(\frac{3}{2}\right)
\Bigg[\delta \tan \beta \frac{3 m_h^2}{s - m_h^2} + \delta^2 \tan \beta \frac{3 m_h^2}{s-m_s^2} \nonumber \\
&&+\frac{\delta \tan \beta}{\sqrt{1-\frac{4 m_h^2}{s}}\sqrt{A}} \ln \left(\frac{m_h^2-\frac{1}{2}\sqrt{A+4s(m_h^2-m_s^2)}
+\frac{1}{2} \sqrt{1-\frac{4 m_h^2}{s}}\sqrt{A}}{m_h^2-\frac{1}{2}\sqrt{A+4s(m_h^2-m_s^2)}
-\frac{1}{2} \sqrt{1-\frac{4 m_h^2}{s}}\sqrt{A}} \right) \nonumber \\
&&+\frac{\delta^2 \tan \beta}{\sqrt{1-\frac{4 m_h^2}{s}}\sqrt{A}} \ln \left(\frac{2m_h^2-m_s^2-\frac{1}{2}\sqrt{A+4s(m_h^2-m_s^2)}
+\frac{1}{2} \sqrt{1-\frac{4 m_h^2}{s}}\sqrt{A}}{2m_h^2-m_s^2-\frac{1}{2}\sqrt{A+4s(m_h^2-m_s^2)}
-\frac{1}{2} \sqrt{1-\frac{4 m_h^2}{s}}\sqrt{A}} \right) \Bigg]\\
&& \bullet~ a_0(ss \rightarrow hs) = - \frac{\lambda_H}{4 \pi}\kappa_{ss \rightarrow hs}\left(\frac{3}{2}\right)
\Bigg[\delta^2 \tan \beta \frac{3 m_h^2}{s - m_h^2} + \delta \tilde{\delta} \tan \beta \frac{3 m_h^2}{s-m_s^2} \nonumber \\
&&+\frac{\delta^2 \tan \beta}{\sqrt{1-\frac{4 m_s^2}{s}}\sqrt{B}} \ln \left(\frac{m_s^2-\frac{1}{2}\sqrt{B+4s(m_s^2-m_h^2)}
+\frac{1}{2} \sqrt{1-\frac{4 m_s^2}{s}}\sqrt{B}}{m_h^2-\frac{1}{2}\sqrt{B+4s(m_s^2-m_h^2)}
-\frac{1}{2} \sqrt{1-\frac{4 m_s^2}{s}}\sqrt{B}} \right) \nonumber \\
&&+\frac{\delta \tilde{\delta} \tan \beta}{\sqrt{1-\frac{4 m_s^2}{s}}\sqrt{B}} \ln \left(\frac{2m_s^2-m_h^2-\frac{1}{2}\sqrt{B+4s(m_s^2-m_h^2)}
+\frac{1}{2} \sqrt{1-\frac{4 m_s^2}{s}}\sqrt{B}}{2m_s^2-m_s^2-\frac{1}{2}\sqrt{B+4s(m_s^2-m_h^2)}
-\frac{1}{2} \sqrt{1-\frac{4 m_s^2}{s}}\sqrt{B}} \right) \Bigg]\nonumber \\
\end{eqnarray}
Note that two new variables, $\hat{\delta}$ and $\delta$, for later works and the simplicity are introduced here and in our main body,
\begin{equation}
\hat{\delta} \equiv \frac{\lambda_{HS}}{\lambda_H}~~,~~\delta \equiv \frac{1}{6}\frac{\lambda_{HS}}{\lambda_H}.
\end{equation}

\newpage
\begin{center}
\acknowledgments
\end{center}
This work was supported in part by the National Research Foundation of Korea (NRF) grant funded by the Korea government of the Ministry of Education, Science and Technology (MEST) (No. 2011-0029758).

\end{document}